\documentclass[a4paper]{article}

\usepackage{graphicx} 

\newcommand\lsim{\mathrel{\rlap{\lower4pt\hbox{\hskip1pt$\sim$}}
\raise1pt\hbox{$<$}}}
\newcommand\gsim{\mathrel{\rlap{\lower4pt\hbox{\hskip1pt$\sim$}}
    \raise1pt\hbox{$>$}}}
\newcommand{\ea}{{\em et al.}\ }

\newcommand{\eref}[1]{(\ref{eq:#1})} 
 
\newcommand{\fref}[1]{Figure \ref{fig:#1}} 
\newcommand{\incfig}[2]{\includegraphics[width=#1\textwidth]
  {#2}}

\newcommand{\pl}{\partial}
 
\begin{document}

\title{On the evolution of abelian-Higgs string networks}

\author{J.N.~Moore and E.P.S.~Shellard \\ \\
  Department of Applied Mathematics and Theoretical Physics,\\
  University of
  Cambridge, \\
  Silver St.,\\
  Cambridge CB3 9EW, U.K.}

\date{\today}

\maketitle

\begin{abstract}
  We study the evolution of abelian-Higgs string networks in numerical
  simulations. These are compared against a modified
  velocity-dependent one scale model for cosmic string network
  evolution.  This incorporates the contributions of loop production,
  massive radiation and friction to the energy loss processes that are
  required for scaling evolution.  We find that the loop distribution
  statistics in the simulations are consistent with the long-time
  scaling of the network being dominated by loop production.  For an
  oscillating sinusoidal perturbation, we also demonstrate that the
  power emitted into massive radiation decays strongly with
  wavelength.  Putting these observations together and extrapolating,
  we believe there is insufficient evidence to reject the the standard
  picture of string network evolution in favour of one where direct
  massive radiation is the dominant decay mechanism, a proposal which
  has attracted much recent interest.
\end{abstract}

\section{Introduction}
\label{sec:Intro}

Vortex-string networks are important in a variety of contexts, whether
in condensed matter physics or cosmology\cite{VilShe}. If we are 
to obtain a quantitative description of these networks, then 
we must also properly understand their `scaling' evolution as well as
the decay mechanisms which maintain it. The abelian-Higgs model, a 
relativistic version of the Ginzburg-Landau theory of superconductors, 
provides a convenient  
testbed for developing detailed models for this evolution. On the one hand, 
the relatively 
simple field theory can be studied directly in three-dimensional simulations.
On the other, a straightforward reduction to a one-dimensional effective 
theory---the Nambu action---can also be studied numerically, though over 
a much wider dynamic range. 

In a cosmological context, a rather simple `one-scale' model of string 
evolution has emerged\cite{Kibble, MarShe} which appears to successfully 
describe
the large-scale features of an evolving string network\cite{BB,AS}, 
though with subtleties 
remaining on smaller scales. In this simple model, the average number of long
strings in a horizon volume remains fixed as it expands, a rapid dilution 
made possible through reconnections resulting in loop production.  The 
loops, in this standard picture, oscillate relativistically and 
decay through gravitational radiation.
The subtlety here concerns small loop creation; gravitational
radiation backreaction effects should act on the long string network 
eliminating small wavelength modes, thus setting a minimum loop 
creation size\cite{BB,ACK}.  This backreaction length 
`scales' with the horizon size (for GUT-scale strings it should be 
approximately $10^{-4}t$), and so loop sizes should also be scale-invariant, 
albeit tiny and, as yet, not adequately 
probed by Nambu string simulations.

Recently this standard picture for network evolution has been questioned
on the basis of abelian-Higgs field theory simulations\cite{VAH}. The 
authors suggest that the primary energy loss mechanism by long strings
is direct massive radiation, rather than loop creation.  This is contrary
to qualitative expectations that the presence of a large mass threshold 
should exponentially suppress massive particle production for any long 
wavelength oscillatory string modes, that is, those much larger than 
the string width\cite{SreThi}. Evidence in support of this
claim rests primarily on the study of large amplitude oscillations 
of a single string and network simulations
in which the loop density is observed to be low. 

The aim of the present work is to consider these
issues by taking a more detailed look at radiation from an 
oscillating string and the available decay mechanisms for 
an evolving network.  In high resolution and low noise simulations
of a perturbed string, we are able first to demonstrate that qualitative 
expectations for massive radiation suppression are correct.
Next, by analytically modelling the small-scale results from 
field theory network simulations,
we are able to argue that these can be sensibly and consistently 
extrapolated to the large lengthscales relevant for Nambu simulations 
and cosmology.  Although complex nonlinear processes are at work on 
small-scales in the field theory simulations, loop and `protoloop' production 
already appears to be the be the dominant energy decay mechanism.

\section{Massive radiation from an oscillating string}
\label{sec:soliton}

As stated above, we shall be considering strings in the Abelian Higgs
model, the simplest producing gauged vortex-lines. The Lagrangian density
is
\begin{equation}
  {\mathcal{L}} = (D_{\mu}\phi)^*(D^{\mu}\phi) 
  - \frac{1}{4}F_{\mu\nu}F^{\mu\nu}
  - \frac{\lambda}{4}(\phi\phi^*-\eta^2)^2.
  \label {eq:ld}
\end{equation}
The covariant derivative $D_\mu$ acts on $\phi$ as $\pl_\mu-ieA_\mu$.
The Higgs field mass is $\sqrt{\lambda}\eta$ and that of the vector
field is $\sqrt{2}e\eta$. After rescaling the only free parameter in
the model is the ratio of these two masses, which we take to be unity
(specifically, $\lambda/2 = \eta = e = 1$).  This corresponds to the
Bogomol'nyi limit for vortices in two dimensions.  The system of
evolution equations resulting from (\eref{ld}) is solved using the
standard lattice gauge theory methods proposed by Myers {\it et al.}\ 
\cite{MMR} (we also always adopt the temporal gauge $A_0=0$).  Initial
data is evolved forward in time using a leap-frog discretization of
Hamilton's equations with a short time-step.  Throughout our
simulations the boundary conditions in use are periodic up to a gauge
transformation in all cases and exactly periodic for network
simulations.

First, we consider a perturbed string lying in the $z$-direction and
oscillating in a periodic box of sidelength $L$.  The starting
configuration for each simulation is based on a simple ansatz for
which the core position varies sinusoidally along the string length
with amplitude $y = A \sin (2\pi z/L)$.  We use profile functions
obtained by solving the one dimensional field equations for a
cylindrical string numerically.  If, as we suggest, the perturbed
string is weakly radiating for large $L$ then it is necessary to
distinguish the 'true' radiation from that due to the spurious modes
introduced by inappropriate initial conditions. We achieve this by
relaxing the gauge links in our lattice system to their energetic
minimum, while fixing the Higgs field, and hence the position of the
string core (this removes the worst modes). Further spurious modes
associatied with the Higgs field modulus are then removed by a
releasing all the fields for a short period to evolve via `gradient
flow', that is, purely first-order dissipative evolution.

\begin{figure}[!htbp]
  \begin{center}
    \leavevmode \incfig{1.0}{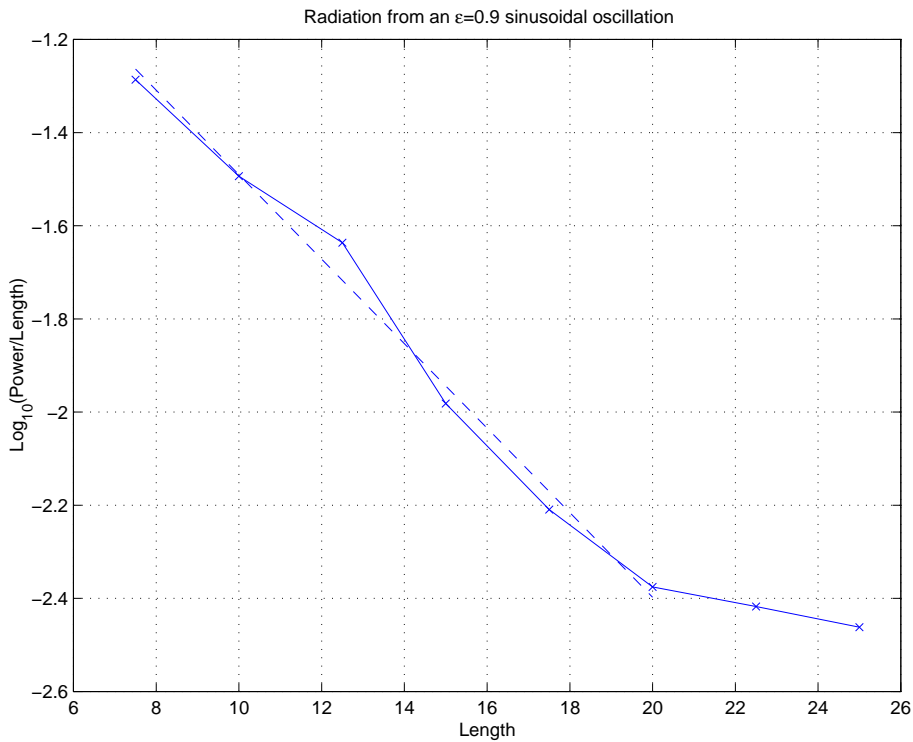}
  \end{center}
  \caption[]{
    The power per unit length emitted by a oscillating 
    strings as a function of length $L$, initially given a fixed 
    relative amplitude $\epsilon = 2\pi A/L = 0.9$.
    The fall-off with increasing length is not consistent with the
    $L^{-1}$ required for scaling. The dashed line is a best
    least-squares fit to exponential fall-off. 
    }
  \label{fig:powlen2}
\end{figure}

To calculate radiative power we look at the rate of change of energy
outside a region of fixed radius around the position of the
unperturbed string. We take the average initial energy increase as the
radiation rate and the effect of varying the string length is
illustrated in \fref{powlen2}. Here, the relative amplitude for the
initial data was kept at a large fixed value, $\epsilon \equiv 2\pi
A/L = 0.9$; it is important that this value is below unity because,
for $\epsilon \gsim 1.0$, the perturbed strings correspond to
degenerate relativistic Nambu configurations which radiate
pathologically \cite{BatShe}.\footnote{For a relative amplitude
  $\epsilon \gsim 1$, a sinusoidal Nambu string develops `lumps', that
  is, finite regions of string which pile up at a point moving at the
  speed of light.  Not surprisingly, radiative backreaction in a field
  theory simulation is severe in such regions, leading to the loss of
  a constant length of string in the first oscillation; initially,
  then, we would have a power per unit length $P \approx L^{-1}$. We
  do not expect such perfectly degenerate string configurations in a
  cosmological context.}  An exponential trend as a function of length
$L$ is apparent, which can be approximately fitted by $P \propto
\exp(-L/L_{\rm m})$ with $L_{\rm m}\approx 4.8$.  For smaller amplitude,
there is an even stronger dependence on $L$, so we can be confident
that $L_{\rm m} \lsim 4.8$ for $\epsilon \lsim 0.9$ (as we describe in
more detail elsewhere \cite{MorShe}).  Certainly the decay of emitted
power between the shortest and longest $L$ shown here is consistent
only with a higher power law than is required for scaling, that is, $P
\propto L^{-n}$ with $\hbox{$n>\!>1$}$.  For $L$ significantly larger
than shown in \fref{powlen2}, the minute power loss becomes
indistinguishable from the background numerical noise inherent in the
simulations.  Note that the actual spectral nature of this radiation
and its relationship to the string perturbation lengthscale $L$
is discussed at length elsewhere \cite{MorShe}.

The strong scale-dependence of massive radiation we observe is not
consistent with the behaviour found in ref.~\cite{VAH}, where they
suggest that the power per unit length scales merely as $P \propto
L^{-1}$. We can see two possible explanations for this clear
discrepancy: First, the results reported in ref.~\cite{VAH} are for
only a single relative perturbation amplitude $\epsilon = \pi$.
This is a strongly nonlinear and degenerate regime for which
pathologically strong radiation is expected.  Secondly, for large
amplitudes the unrelaxed initial ansatz in ref.~\cite{VAH} is
inaccurate, particularly for the gauge fields.  As we have observed in
our own simulations, this can lead to spurious modes from the initial
relaxation which swamp the radiation due to the string oscillations.

\section{String network evolution}
\label{sec:network}

We have performed network simulations using periodic cubic lattices of
side length 250 and greater. The principal physical parameter in each
simulation is the initial correlation length. To establish a suitable
network configuration we take an initial configuration with a flat
initial power spectrum of fluctuations in the Higgs field, centered
around zero, and zero gauge field. This is evolved forward in time
with Hamiltonian dynamics to increase the correlation length and then
acted upon by a short period of gradient flow evolution to reduce the
initial inter-string energy density to a negligible value, since the
one-scale model requires that this be zero initially. Subsequently the
simulation is evolved using Hamiltonian dynamics. The Gauss constraint
is satisfied initially as we start from rest, and is preserved by the
equations of motion.

The time-step is chosen to be sufficiently small to allow the
Hamiltonian to be conserved within 1 per cent over the course of a
run. The choice of spatial lattice spacing is constrained by the
desire to meet two conflicting criteria: (i) to simulate a volume
which is orders of magnitude larger than the string width, and (ii) to
accurately represent the continuum theory. Throughout we have chosen a
lattice size of 0.5, which is as large as we feel is reasonable given
that at larger spacings there is a significant potential barrier
associated with the lattice and that for oscillating strings of length
$\approx 15$ we have observed greatly increased radiation using a larger
lattice spacing.

To characterise the network configuration at a given point in time we
assign positions of zeros of the Higgs field to lattice plaquettes
according to the winding of the Higgs phase around each
plaquette. This allows us to calculate
correlation lengths and loop distribution statistics \cite{AS}. When combined
with similar data from a nearby time-step we can also estimate the
velocity of each string segment. The procedure for calculating velocities 
is relatively
vulnerable to numerical errors due to uncertainty in the `true'
position of the string network and the sensitivity of measured
velocities to this.

In \fref{length} we see general consistency with the results of Figure
1 of Vincent \ea. In both cases the network correlation length grows
approximately linearly throughout the simulations. The most notable
deviations from linearity is seen in the run at largest lattice
spacings of ref.~\cite{VAH}, where the rate of growth of the
correlation length with time is $\sim 0.5$, some 50 per cent greater than
in the rest of the simulations. We speculate that this increased rate
of growth is caused by lattice effects. The initial brief burst of
growth in correlation length seen by Vincent \ea in several
simulations is not seen in our simulations and may arise from the
slightly different procedures used to set up the initial network
configurations.

\begin{figure}[!htbp]
  \begin{center}
    \leavevmode \incfig{1.0}{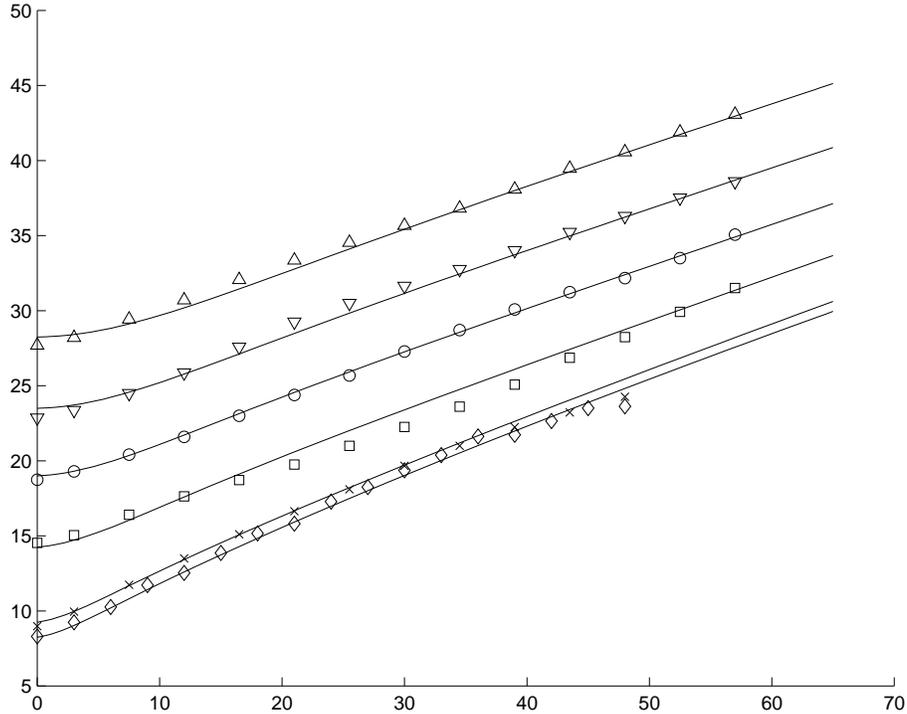}
  \end{center}
  \caption[]{
    The spatial correlation length $L$ as a function of time
    for a series of simulations with a grid size of
    $250^3$. 
    A velocity dependent `one-scale' model is used to fit the data (solid
    lines).
    }
  \label{fig:length}
\end{figure}
\begin{figure}[!htbp]
  \begin{center}
    \leavevmode \incfig{1.0}{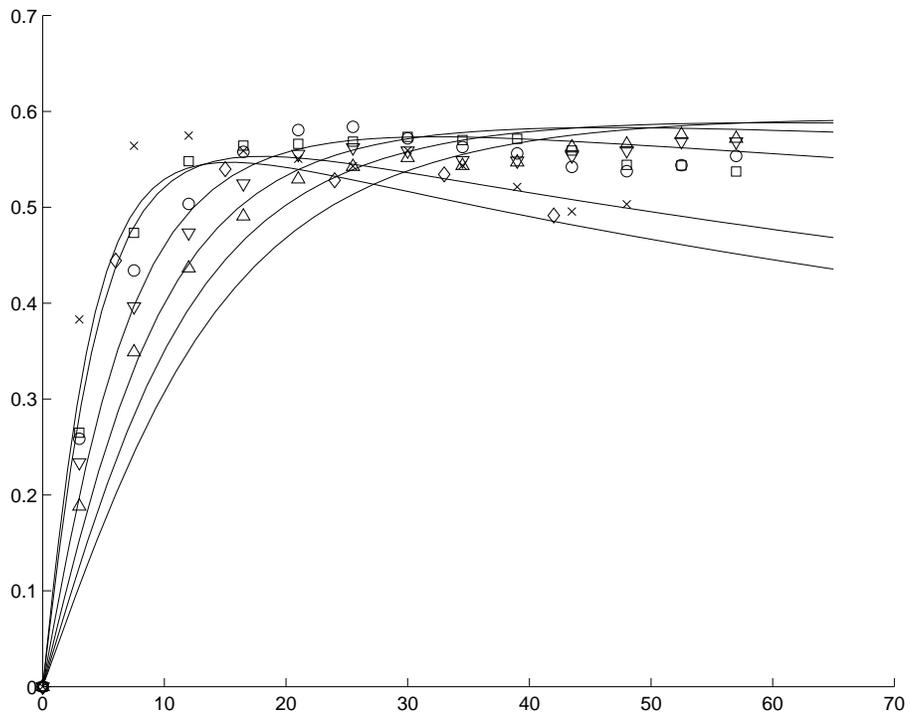}
  \end{center}
  \caption[]{The average rms velocity $v$ as a function of time
    for a series of simulations with a grid size of $250^3$. 
    The solid lines indicate the fit of our simple analytic model.}
  \label{fig:velocity}
\end{figure}

\section{Analytic modelling of network evolution}
\label{sec:analytic}

If we regard the one-dimensional Nambu action as providing a
satisfactory first approximation to string evolution then we can
employ a velocity-dependent `one-scale' model to describe the
evolution of a string network \cite{MarShe}.  The long string network
energy density $\rho_\infty$ is susceptible to three possible energy
loss mechanisms, friction, loop production and direct massive
radiation which are phenomenologically summarized in the following
averaged evolution equation \cite{MarShe}:
\begin{equation}
  \frac{d\rho_\infty}{dt} = \frac{v^2 \rho_\infty}{L_{\rm f}}+ 
  \frac{\bar c v \rho_\infty }{L} + {\rho_\infty} f
  ({L}/{L_{\rm m}})\,,
  \label {eq:marshe0}
\end{equation}
where $L_f$ is the friction length, $\bar c$ is the loop chopping 
efficiency, and $f({L}/{L_{\rm m}})$ is proportional to the power per 
unit length of massive radiation.  Employing the definition for the 
correlation length $\rho_\infty \equiv \mu/L^2$ and noting a supplementary 
equation for the rms velocity $v$ we obtain \cite{MarShe}
\begin{equation}
  \frac{dL}{dt} = \frac{Lv^2}{2L_{\rm f}}+ \frac{\bar c v}{2} + \frac{L}{2}f
  ({L}/{L_{\rm m}})\,,
  \label {eq:marshe1}
\end{equation}
\begin{equation}
  \frac{dv}{dt} = (1-v^2)\left[\frac{1-2v^2}{L} - \frac{v}{L_{\rm f}}
    - a\frac{2\bar c v^2}{L} \right]\,.
  \label {eq:marshe2}
\end{equation}
We provide some explanation for the origin of these terms below, but
note that the first term in the velocity equation is due to the
acceleration of strings with a typical curvature radius $L$ (with
$v\rightarrow 1/\sqrt2$ for free relativistic motion).  Given the
difficulties in measuring and normalizing 
this velocity in the field theory simulations, we also allow 
for lattice
discretization effects through the phenomenological 
parameter $a$; this term takes a
form which could also incorporate momentum losses due to loop production. 

A potentially important energy loss mechanism for the string network
is friction through interactions with background energy density $\bar
\rho$.  For the simulations described here with dissipative initial
conditions, this background arises dynamically through the decay of
the string network into massive particles, that is, $\bar\rho =
\rho_{\infty,\rm i} - \rho_\infty$.  Characterising this in terms of a
friction lengthscale above which friction dominates, we have $L_{\rm
  f}\equiv \mu\beta/{\bar\rho} = \beta(1/L_{\rm i}^2-1/L^2)^{-1}$ (for
$L\rightarrow \infty$ in flat space, $L_{\rm f} \rightarrow \beta
L_{\rm i}^2$.  We can bound the friction coefficient $\beta$ by
beginning with a high density string network and assuming a late-time
friction-dominated regime.  An example is the lowest curve in
\fref{length}, where friction affects the scaling behaviour, most
obviously through the velocity in \fref{velocity}; this provides a
lower limit of $\beta \gsim 0.2$.  From this we can infer that
friction provides less than 10\% of the energy losses for the duration
of the simulations in \fref{length}, if the initial correlation length
satisfies $L_{\rm i} \gsim 20$.  Hence, assuming friction to be
insignificant for the lowest density simulations, we can make a more
accurate estimate $\beta \approx 0.3$.

Loop production is caused by the intersections and self-intersections
of long strings. For the simulations least affected by friction, let
us assume the standard picture with energy losses predominantly due to
this loop production.  In this case, from a fit to the correlation
length and velocities we can obtain the asymptotic value $\bar
c\approx 0.75$ (with some small momentum losses $a\approx 0.45$).  This
value of $\bar c$ is consistent with previous determinations of the
loop chopping efficiency in flat space simulations \cite{VHS}; note
that the velocity-dependence in our definition implies that their $c =
\bar c/v$).

Direct massive radiation will provide some  energy losses
in the early evolution, but this should diminish as the correlation
length increases, consistent with the behaviour of the
oscillating string discussed in section 2. From \fref{length} we can
see that there is considerable evidence for an additional
scale-dependent initial energy loss mechanism, since the initial
slopes for the smaller correlation lengths are clearly steeper than
those with $L_{\rm i} \gsim 15$. From a fit to \fref{length}, the
function describing the energy losses into massive radiation should
behave approximately as $f(L/L_{\rm m}) \approx \gamma \exp(- L/L_{\rm
  m})$ with $L_{\rm m}\sim 4.8$ for the present parameters.  The overall
strength $\gamma$ can only be rougly estimated from the initial slopes
in \fref{length} to yield $\gamma\sim 0.15$.  This implies, consistent
with \fref{length}, that massive radiation provides a strong initial
contribution only for $L \lsim 10$ and, subsequently, it is rapidly
curtailed.

What we obtain finally is the self-consistent fit of our analytic
model to the correlation length and velocity evolution shown in
\fref{length}.  Asymptotically this is quantitatively in agreement
with the standard picture of long string evolution, and it also
reproduces predicted qualitative features.  However, if we choose to
ignore the loop contribution as in ref.~\cite{VAH} and we assume that
massive radiation losses are scale-invariant ($f\propto L^{-1}$), then
it is also possible to fit the asymptotic data with this analytic
model, although the initial qualitative features would not be as consistent.
 The question of whether the dominant loop loss
interpretation is correct, therefore, must hinge on more precise
information about the loop distribution and how it evolves in time.

\section{Competing energy loss mechanisms}
\label{sec:competing}

In using these simulations to determine the dominant decay mechanisms
for a cosmic string network we face a rather obvious difficulty.  If
the standard cosmological picture of loop production is correct, then
we know from high resolution Nambu string simulations that the typical
loop creation scale $\bar\ell$ relative to the horizon is $\alpha
\equiv \bar\ell/t \lsim 10^{-3}$, while radiation backreaction
estimates suggest that $\alpha \sim 10^{-4}$ for GUT-scale strings.
The present field theory simulations, however, have a dynamic range
which is at least two orders of magnitude poorer, so we cannot
realistically hope to probe these small-scale regimes. In consequence,
we might be surprised to be able to identify any loops at all and we
certainly cannot expect their average creation size $\bar\ell$ to
`scale' relative to $L$.  As we observe numerically, most loops which
are created have radii comparable to the string thickness, with $\bar
\ell \sim \pi$ almost constant throughout.  Indeed, simulation
visualisations indicate that most energy is lost via `protoloops',
that is, small-scale highly nonlinear, but coherent, regions of energy
density (as illustrated in \fref{protoloop}).  Consistent with the
Nambu simulations, this `protoloop' production occurs---like small
loop formation---in high curvature regions where the strings collapse
and become very convoluted \cite{SheAll}.

\begin{figure}[!htbp]
  \begin{center}
    \leavevmode \incfig{1.0}{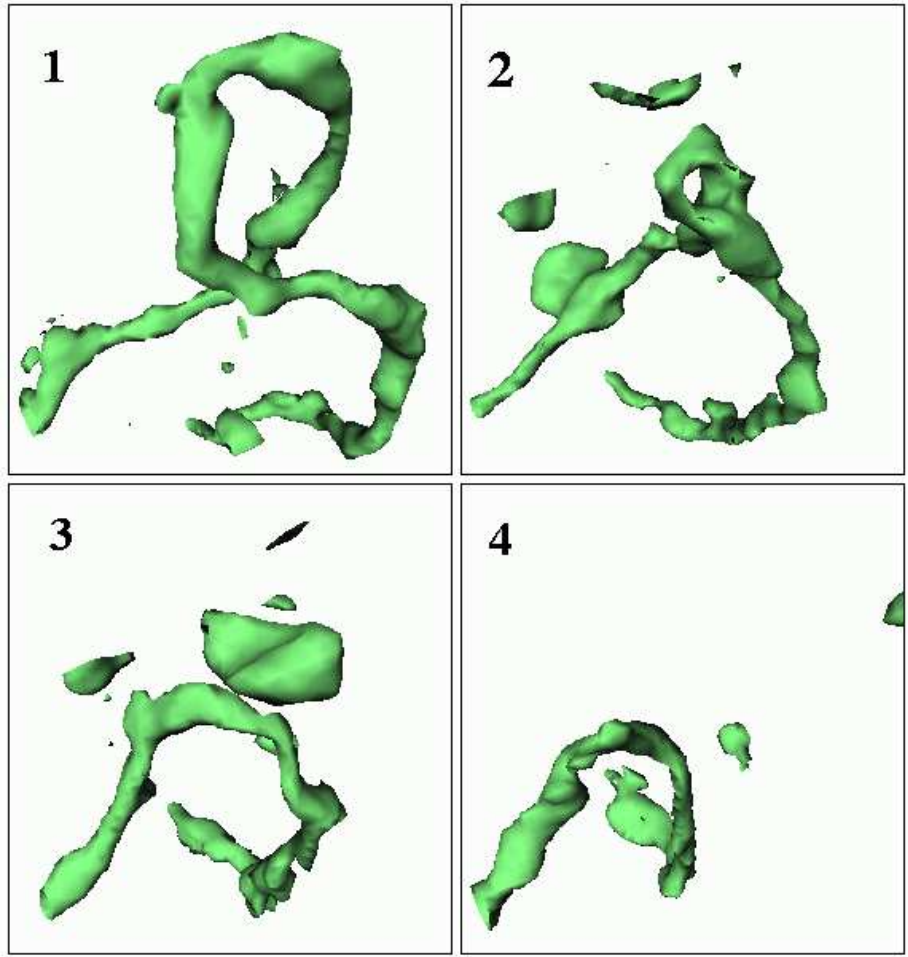}
  \end{center}
  \caption[]{
    Four stages of the formation of  `protoloops' in the field theory
    simulations. :  A highly curved region of string (i) collapses to 
    form nonlinear `lumps' in the energy density (ii) which are lost
    (iii,iv) when the
    the string's topological winding intercommutes or 
    annihilates in this region.  Energy density contours are plotted.
    }
  \label{fig:protoloop}
\end{figure}

A proportion of these `protoloops' have sufficient topology to be
identified as loops by our numerical diagnostics, so we can estimate
the energy loss via this pathway relative to other mechanisms. Because
of dynamic range limitations, all small loop trajectories in these
simulations are self-intersecting and loops decay almost as fast as
allowed by causality. Their time-dependent length, therefore, can be
given by $\ell \approx \ell_{\rm c} - 4(t-t_{\rm c})$ in the range
$0< t-t_{\rm c} < \ell_{\rm c/4}$ with the loop creation length
$\ell_{\rm c}$ and time $t_{\rm c}$.  Hence, at any one time, the
string energy loss through loop decay can be simply approximated by
$\dot\rho_\ell \approx - 4\mu n_\ell$, where $n_\ell$ is the loop
number density.\footnote{Note that this is quite different to the
  cosmological Nambu limit where loops can be non-intersecting and
  long-lived.}  For linear scaling to pertain, a dominant energy loss
mechanism must behave as $\dot \rho \propto t^{-3}$, implying $n
\propto t^{-3}$ so the loop energy density would behave as $\rho_\ell = \mu
\bar \ell n_\ell \propto \rho_\infty/t$ (with $\bar \ell$ constant).
Thus, contrary to one of the conclusions in ref.~\cite{VAH}, loops can
be an important decay mechanism even though their relative
contribution to the total string energy density is falling rapidly as
$t^{-1}$.

\fref{relenergy} illustrates the relative small loop contribution to
the overall energy density losses throughout the particular simulation
beginning with $L_{\rm i} = 15$.  This is compared with the
estimated analytic fit for friction, loop losses and massive
radiation.  We can observe that the measured loop energy losses grow
steadily towards the analytic loop contribution, which for these
simulations we assume must include `protoloops' both with topology and
without it.  We can see that the proportion of loops---the topological
`protoloops'---gradually grows to meet or even overtake the analytic loop 
contribution by the end of the
simulation. In fact our analytic calculation deserves closer quantitative
scrutiny because it is a significant 
overestimate; the histogram plotted in \fref{relenergy} only gives 
$\mu n_{\ell} L^3$ rather than four times this quantity. 

With this clear qualitative trend evident, it is not unreasonable to 
conjecture that an extrapolation by
another twenty orders of magnitude to cosmological scales will
imply that the loop contribution will be completely
dominant.  As the typical string perturbation lengthscale grows and is
affected by radiative backreaction, it is again reasonable to suppose
that the typical loop creation size will also grow, becoming many
orders of magnitude larger than the string thickness.  We conclude
that these field theory simulations, once we account for their small
dynamic range, are consistent with the standard picture of long string
network evolution via small loop production.  At the very least, the
simulations do {\it not} provide compelling evidence that cosmological
strings will decay primarily through the direct radiation of
ultra-massive particles.

\begin{figure}[!htbp]
  \begin{center}
    \leavevmode \incfig{1.0}{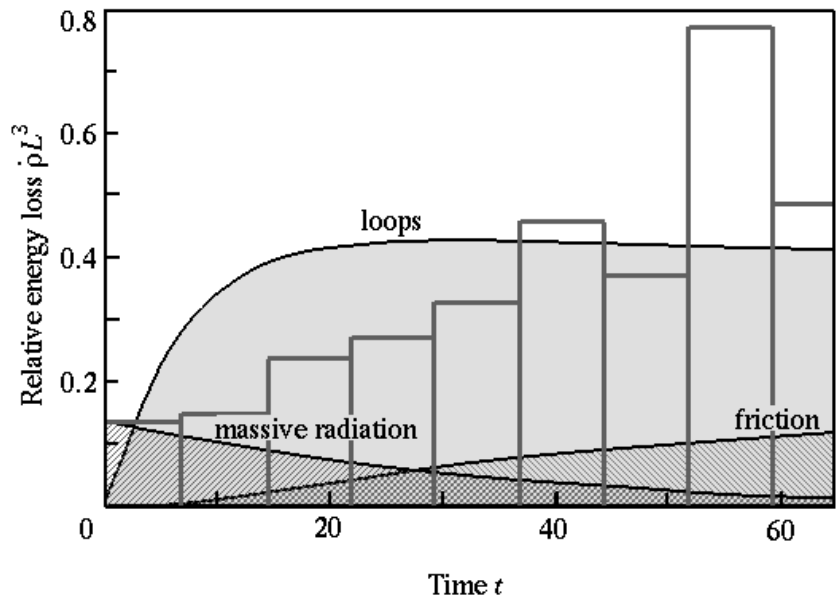}
  \end{center}
  \caption[]{The relative contributions of the different components
    to the overall energy density losses of the string network, as
    predicted by the analytic model for the simulation in
    \fref{length} with $L_{\rm i} = 15$. Superposed onto this figure
    is a histogram of energy loss into loop production estimated directly
    from measurements of the time-varying loop density in this 
   particular simulation. 
        It is apparent that the loop energy loss contribution is becoming 
        more important as the simulation progresses.}
  \label{fig:relenergy}
\end{figure}

\section{Conclusion}
\label{sec:conclusion}

We conclude from these numerical results and their analytic interpretation
that the standard picture provides a more 
coherent and adequate model for string network evolution 
than the more radical alternative based
on direct massive radiation\cite{VAH}.  However, this is not to suggest 
that we have provided
a complete or detailed description of the complex nonlinear processes that underlie 
network evolution on these small scales.  Rather, first, 
we have shown qualitatively 
that loop production is important for network evolution and should become
more so when extrapolated to large scales.  Secondly, we have also 
demonstrated that massive radiation is strongly suppressed for 
long wavelength modes, implying that it is an inadequate decay 
channel for maintaining network `scaling'.  We are currently investigating 
both these aspects in more quantitative detail\cite{MorShe}. 

These results have significant cosmological implications.  Our
expectation is that massive particles will only be produced
infrequently in highly nonlinear string regions, such as at cusps and
reconnections.  The ensuing flux of cosmic rays should be relatively
low\cite{BR,PS}.  The more recent estimates of the ensuing cosmic ray
flux from direct massive radiation from strings appear to be overly
optimistic\cite{VAH,Bha}.  This work points to the need for caution in
making cosmological extrapolations from small-scale numerical
simulations and to the need for further progress understanding string
radiation backreaction.

\section*{Acknowledgements}

We are grateful for useful discussions with Richard Battye, Mark
Hindmarsh and Graham Vincent.  The simulations were performed on the
COSMOS Origin2000 supercomputer which is supported by Silicon
Graphics, HEFCE and PPARC.


\def\hang{\noindent}

\def\jnl#1#2#3#4#5#6{\hang{#1, {\it #4\/} {\bf #5}, #6 (#2).}}


\def\jnlerr#1#2#3#4#5#6#7#8{\hang{#1, {\it #4\/} {\bf #5}, #6 (#2).
    {Erratum:} {\it #4\/} {\bf #7}, #8.}}


\def\jnltwo#1#2#3#4#5#6#7#8#9{\hang{#1, {\it #4\/} {\bf #5}, #6 (#2);
    {\it #7\/} {\bf #8}, #9.}}

\def\prep#1#2#3#4{\hang{#1 [#2],  #4.}}

\def\myprep#1#2#3#4{\hang{#1 [#2], '#3', #4.}}

\def\proc#1#2#3#4#5#6{\hang{#1 [#2], `#3', in {\it #4\/}, #5, eds.\ (#6).}
  }
\def\procu#1#2#3#4#5#6{\hang{#1 [#2], in {\it #4\/}, #5, ed.\ (#6).}
  }

\def\book#1#2#3#4{\hang{#1 [#2], {\it #3\/} (#4).}
  }

\def\genref#1#2#3{\hang{#1 [#2], #3}
  }


\def\prl{Phys.\ Rev.\ Lett.}
\def\pr{Phys.\ Rev.}
\def\pl{Phys.\ Lett.}
\def\np{Nucl.\ Phys.}
\def\prp{Phys.\ Rep.}
\def\rmp{Rev.\ Mod.\ Phys.}
\def\cmp{Comm.\ Math.\ Phys.}
\def\mpl{Mod.\ Phys.\ Lett.}
\def\apj{Ap.\ J.}
\def\apjl{Ap.\ J.\ Lett.}
\def\aap{Astron.\ Ap.}
\def\cqg{Class.\ Quant.\ Grav.} 
\def\grg{Gen.\ Rel.\ Grav.}
\def\mn{M.$\,$N.$\,$R.$\,$A.$\,$S.}
\def\ptp{Prog.\ Theor.\ Phys.}
\def\jetp{Sov.\ Phys.\ JETP}
\def\jetpl{JETP Lett.}
\def\jmp{J.\ Math.\ Phys.}
\def\cupress{Cambridge University Press}
\def\pup{Princeton University Press}
\def\wss{World Scientific, Singapore}

\bibliographystyle{hunsrt}
\bibliography{../bibs/standard}

\end{document}